\documentclass[twocolumn]{aastex631}
\shorttitle{EUV jet driven by transition region micro-jets}
\shortauthors{Wei et al.}
\graphicspath{{./}{figures/}}

\begin{document}

\title{An EUV jet driven by a series of transition region micro-jets}

\correspondingauthor{Zhenghua Huang}
\email{z.huang@sdu.edu.cn}

\author[0000-0002-0210-6365]{Hengyuan Wei}
\affiliation{Shandong Key Laboratory of Optical Astronomy and Solar-Terrestrial Environment, Institute of Space Sciences, Shandong University, 264209, Weihai, Shandong, China}

\author[0000-0002-2358-5377]{Zhenghua Huang}
\affiliation{Shandong Key Laboratory of Optical Astronomy and Solar-Terrestrial Environment, Institute of Space Sciences, Shandong University, 264209, Weihai, Shandong, China}

\author[0000-0002-8827-9311]{Hui Fu}
\affiliation{Shandong Key Laboratory of Optical Astronomy and Solar-Terrestrial Environment, Institute of Space Sciences, Shandong University, 264209, Weihai, Shandong, China}

\author{Ming Xiong}
\affiliation{State Key Laboratory of Space Weather, National Space Science Center, Chinese Academy of Sciences, Beijing, China}
\affiliation{College of Earth and Planetary Sciences, University of Chinese Academy of Sciences, Beijing 100049, China}

\author[0000-0001-8938-1038]{Lidong Xia}
\affiliation{Shandong Key Laboratory of Optical Astronomy and Solar-Terrestrial Environment, Institute of Space Sciences, Shandong University, 264209, Weihai, Shandong, China}

\author{Chao Zhang}
\affiliation{Shandong Key Laboratory of Optical Astronomy and Solar-Terrestrial Environment, Institute of Space Sciences, Shandong University, 264209, Weihai, Shandong, China}

\author{Kaiwen Deng}
\affiliation{Shandong Key Laboratory of Optical Astronomy and Solar-Terrestrial Environment, Institute of Space Sciences, Shandong University, 264209, Weihai, Shandong, China}

\author{Haiyi Li}
\affiliation{Shandong Key Laboratory of Optical Astronomy and Solar-Terrestrial Environment, Institute of Space Sciences, Shandong University, 264209, Weihai, Shandong, China}



\begin{abstract}
Jets are one of the most common eruptive events in the solar atmosphere, 
and they are believed to be important in the context of coronal heating and solar wind acceleration.
We present an observational study on a sequence of jets with the data acquired with the Solar Dynamics Observatory (SDO) and the Interface Region Imaging Spectrograph (IRIS).
This sequence is peculiar in that an EUV jet, $\sim29\arcsec$ long and with a dome-like base, appears to be a consequence of a series of transition region (TR) micro-jets that are a few arcsecs in length.
We find that the occurrence of any TR micro-jets is always associated with the change of geometry of micro-loops at the footpoints of the microjets. 
A bundle of TR flux ropes is seen to link a TR micro-jet to the dome-like structure at the base of the EUV jet. 
This bundle rises as a response to the TR micro-jets, with the rising motion eventually triggering the EUV jet.
We propose a scenario involving a set of magnetic reconnections, in which the series of TR micro-jets are associated with the processes to remove the constraints to the TR flux ropes and thus allow them to rise and trigger the EUV jet.
Our study demonstrates that small-scale dynamics in the lower solar atmosphere are crucial in understanding the energy and mass connection between the corona and the solar lower atmosphere, even though many of them might not pump mass and energy to the corona directly.
\end{abstract}

\keywords{Sun: atmosphere---Sun: transition region---Sun: corona---Sun: jet---magnetic reconnection---methods: data analysis}


\section{Introduction} \label{sec:intro}

Jets are one of the most common features in the solar atmosphere, and are thought to be essential phenomena to understand the problems of coronal heating and solar wind acceleration\,\citep{2016SSRv..201....1R}. They are widely observed in the solar atmosphere\,\citep[][]{2021RSPSA.47700217S}, with sizes ranging from several to several-hundred Mega-meter (Mm).
In the corona, jets are transient plasma ejection aligned with magnetic field lines such as open fields or large loops.
There are two types of coronal jets, straight anemone jets and two-side-loop jets\,\citep{1994xspy.conf...29S}. 
They have been observed in X-rays and almost all extreme-ultraviolet(EUV) wavebands\,\citep[e.g.][]{1992PASJ...44L.173S,1996PASJ...48..123S,2009SoPh..259...87N,2018ApJ...854...80H}, and this demonstrates their multi-thermal nature.
Observations have also shown that the hot components of a coronal jet are associated with heating by magnetic reconnection between local closed and open magnetic field, while the cool components are from the small eruptive filaments under jet's base arch \citep{2012ApJ...745..164S}.
The impulsiveness and energetics of a jet could be predicted through its surrounding magnetic field configuration via a model given by \citet{2015A&A...573A.130P}.

\par
Straight anemone jets normally appear as a cusp shape \citep{2011ApJ...735L..18L}, fan-spine structure \citep{2016ApJ...819L...3Z}, inverted-Y \citep{2019ApJ...883..115N} or quasi-circular ribbon \citep{2015PASJ...67...78J}. 
Magnetic structures of the footpoints of straight anemone jets are normally mixed polarities\,\citep{2021RSPSA.47700217S}.
Using the latest Solar Orbiter observations, \citet{2021ApJ...918L..20H} point out that the footpoints of most coronal micro jets occurring at the edge of networks are related to concentrations of magnetic flux and mixed-polarities.
\citet{2010ApJ...714.1762P} show that the magnetic characteristics of a jet could determine its dynamics, such as twist, size, symmetry and null points.
\citet{2016ApJ...820...77W} demonstrate that the dynamics of a jet is sensitive to the ratio of the width of its source region to the footpoint separation of the enveloping loops. 

\par
Straight anemone coronal jets are thought to be produced by magnetic reconnection between emerging bipoles and local open fields \citep{1995Natur.375...42Y,2011ApJ...735L..43S,2012A&A...548A..62H,2015ApJ...804...69Y}.
Both chromospheric surge and X-ray jet can be generated simultaneously from the same magnetic reconnection between rising loops and local open fields that also releases enough energy to produce micro-flares\,\citep{1995Natur.375...42Y}.
When reconnection occurs between local close field and open field, magnetic field lines reconnect to the ambient ones, and form a jet whose base appears to be a structure like an anemone\,\citep{1994ApJ...431L..51S,2015ApJ...815...71C}.
Based on magnetic extrapolation and observing characteristics from EUV, UV, X-ray data, a typical magnetic topology of a jet has been found to include a three-dimensional null point that is associated with an embedded dipole and shows as dome-like or inverted-Y shape \,\citep[e.g.][]{1990ApJ...350..672L,2001ApJ...554..451F,2009ApJ...691...61P,2013ApJ...771...20M,2011ApJ...735L..18L,2012ApJ...746...19Z,2015A&A...573A.130P,2017Natur.544..452W}.
Dome could be destructed by magnetic reconnections between emerging flux ropes and the the fan-spine dome \citep{2015PASJ...67...78J}.

\par
In observations of most straight anemone jets, the rising and eruptions of mini-filaments below the dome seem to be essential\,\citep{2015Natur.523..437S,2017ApJ...835...35H,2017ApJ...842L..20L,2018Ap&SS.363...26L,2018ApJ...859..122L,2019ApJ...872...87L,2017ApJ...851...30X}. 
Mini-filament could be formed due to strong sheared magnetic field, and decrease in the constrain force can lead to its rising and thus results in reconnection with local open field and produces jets\,\citep{2017Natur.544..452W}. 
\citet{2018ApJ...854..155K} suggest that the free energy supplied to jets comes from magnetic shear motion at the polarity inversion lines, and these activities are consistent with a breakout model. 
By studying pseudostreamer jets and CMEs, \citet{2021ApJ...907...41K} find that the determining factor to the dynamic behaviour of an eruption is the ratio of the magnetic free energy related to the filament channel compared to the energy of the overlying flux both inside and outside the pseudostreamer dome.
\citet{2016ApJ...827....4W} suggest that the filamentary structure caused by repeated tearing-like instability in magnetic reconnection provides explanation for intermittent outflows and brightening blobs in jets that have been reported by many studies\,\citep[e.g.][]{2014A&A...567A..11Z,2022FrASS...8..238C}.
\citet{2019ApJ...870..113Z} further confirm that blobs can originate from the tearing instability occurring at the base of a coronal jet or inside a coronal jet.
A recent simulation based on a spheromak configuration demonstrates that magnetic reconnection at the bottom of the spheromak due to emerging processes is the reason of the onset of a coronal jet\,\citep{2021PhPl...28a2901L}.

\par
\citet{2008ApJ...683L..83N} find evidences of propagating Alfv\'en waves along the jet, which are produced by magnetic reconnection in transition region or upper chromosphere. 
\citet{2015ApJ...806...11M} suggested that the driver of a jet can be magnetic-untwisting waves,
which are large amplitude torsional Alfv\'en waves generated in magnetic reconnection and that may dissipate in the corona.
In a dome model, when stress is applied at the photosphere, a 3D null point topology will be driven by continuously twisting motion, and the helical rotating current sheet located along the fan between the sheared spines can naturally produce jets\,\citep{2010ApJ...714.1762P}.
MHD waves, non-thermal electrons, or conduction front associated with a coronal jet can heat the far end of the coronal loop along which the ejecting plasma is flowing\,\citep[e.g.][]{2007PASJ...59S.745S,2020ApJ...897..113H}.

\par
Coronal jets might occur {sequentially} or with multiple phases.
\citet{2021A&A...647A.113Z} report a flare-related coronal jet and find its slow rising phase is caused by reconnection at breakout current sheet and its fast rising phase is caused by reconnection at flare current sheet. 
\citet{2012ApJ...745..164S} report a simultaneously produced bubble-like CME and jet-like CME with two steps including reconnection between the open field and the filament channel and that under the filament channel. 
\citet{2021ApJ...912L..15T} find that the open field reconnected in the occurrence of the first jet sweeps and reconnects with other coronal bright point and thus produces the second jet.
\citet{2017ApJ...844...28S} suggest that the rising mini-filaments reconnect with the dome at different positions several times and this produces multiple jets in one dome system.

\par
Although coronal jets have been studied for decades and many key aspects have been understood, 
their connection to abundant small-scale activities in the lower solar atmosphere has been poorly investigated.
In this paper, we report on multi-wavelength imaging and spectroscopic observations of a sequence of jets including an EUV jet, a series of transition region (TR) micro-jets and a bundle of TR flux ropes.
We will investigate in-depth the evolution of these jets and demonstrate what roles those TR micro-jets play in driving the relatively large EUV jet.
In what follows, we give the data description in Section\,\ref{sec:obs}, results in Section\,\ref{sec:results}, discussion in Section\,\ref{sect:dis} and conclusions in Section\,\ref{sec:conc}.

\section{Observations}
\label{sec:obs}
Our data were obtained on 2016 June 2 from 09:16:39 UT to 10:28:55 UT with a target of AR 12551, by the space-borne Atmospheric Imaging Assembly \citep[AIA,][]{2012SoPh..275...17L} and the Helioseismic and Magnetic Imagers \citep[HMI,][]{2012SoPh..275..207S} on board the Solar Dynamics Observatory \citep[SDO,][]{2012SoPh..275....3P} and the Interface Region Imaging Spectrograph \citep[IRIS,][]{2014SoPh..289.2733D}.
The data analysed in this study mainly include the line-of-sight magnetograms taken with HMI, images with AIA at EUV passbands of 94\,\AA, 171\,\AA, 304\,\AA\ and UV passband of 1600\,\AA, images with IRIS slit-jaw (SJ) Imager at the passbands of 1400\,\AA\ and 2796\,\AA\  and spectral data with IRIS spectrograph. 

\par
The pixel size of the HMI magnetograms is 0.6\arcsec, and the cadence is 45\,s.
The AIA data have a pixel size of 0.6\arcsec\ and a cadence of 12s for EUV images and 24s for images at 1600\,\AA\ passband. 
The IRIS SJ images have a 0.17\arcsec\ pixel size and 18\,s cadence. 
Both HMI and AIA data have been reduced by the standard procedure of {\it aia\_prep.pro} in {\it }solarsoft.
The level 2 IRIS SJ data are used, and no further calibration is required. 

\par
The IRIS spectrograph was running in a ``sit and stare'' mode with 9\,s exposure time from 09:16:39\,UT to 10:29:04\,UT.
In total, it obtained 468 exposures with a cadence of 9.3\,s. 
The spectrograph slit has a width of 0.35\arcsec\ and the data downloaded here cover 119\arcsec\ along the slit.
The relative location of the spectrograph slit is shown in Figure\,\ref{fig:overview}(a) as the dark vertical line.
In this study, we analyse the spectral data of Mg\,{\sc ii} ($1.4\times10^4$\,K), C\,{\sc ii} ($2.5\times10^4$\,K), Si\,{\sc iv} ($7.9\times10^4$\,K) and O\,{\sc iv} ($1.4\times10^5$\,K).
The pixel sampling of wavelength is about 25\,m\AA/pixel.
The production of the spectral data is level 2, from which we have applied the radiometry calibration by following the procedures given in the user's guide provided on the IRIS official website.

\par
The images from different passbands and/or instruments have been co-aligned by using multiple reference structures (such as dark and bright points) in passbands with the closest representative temperatures.

\section{Results}
\label{sec:results}
The sequence of the events occurs near the west limb from 09:16 to 09:50 UT (see Figure\,\ref{fig:overview} and the associated animation).
They consist of a series of micro-jets and a large jet.
The large jet with an inverted-Y shape originates from a dome-like feature (pointed by the white arrow in Figure\,\ref{fig:overview}(a)).
Please note that the dome-like feature is only visible when the jet occurs.
The large jet can be seen at coronal as well as transition region temperatures (see the elongated feature denoted by cyan arrows in Figure\,\ref{fig:overview}).
The series of micro-jets originate from a block of bright points toward the north of the dome-like feature, and in contrast to the large jet they can be seen only in the IRIS SJ 1400\,\AA (i.e. transition region).
In Table\,\ref{tab:jetpara}, we list a number of key parameters of the large jet and these micro-jets.
By investigating these observations in-depth, we suggest that these micro-jets modify the magnetic topology that drives the large jet.
More details are given in the following.

\begin{table*}[!ht]
    \centering
    \caption{Physical parameters of the sequences of jets.}
    \label{tab:jetpara}
    \begin{tabular}{c c c c c c c c c }
        \hline\hline
         Jet&start time&location\tablenotemark{a}&length\tablenotemark{b}&width\tablenotemark{b}&speed\tablenotemark{c}&lifetime&density&passbands\tablenotemark{d}\\
         &(UT)&(X\arcsec,Y\arcsec)&(\arcsec)&(\arcsec)&(km/s)&(s)&($10^{11}$cm$^{-3}$)&(\AA)\\
         \hline
         Large jet&09:38:03&917.1,85.6&29$\pm$0.34&4$\pm$0.34&175$\pm$35&600$\pm$18&$1.5\pm0.3$&1400+EUV\\
         1st micro-jet&09:17:17&917.5,90.0&4.6$\pm$0.34&0.7$\pm$0.34&47$\pm$10&90$\pm$18&--&1400\\
         2nd micro-jet&09:19:08&916.5,93.1&3.3$\pm$0.34&0.8$\pm$0.34&59$\pm$22&54$\pm$18&--&1400\\
         3rd micro-jet&09:22:39&916.4,92.8&6.3$\pm$0.34&0.7$\pm$0.34&50$\pm9$&90$\pm$18&--&1400\\
         4th micro-jet&09:26:39&917.6,90.9&8.0$\pm$0.34&0.6$\pm$0.34&64$\pm$11&70$\pm$18&--&1400\\
         5th micro-jet&09:30:15&917.6,89.9&3.0$\pm$0.34&0.6$\pm$0.34&67$\pm$44&$>$144&--&1400\\
         6th micro-jet&09:32:39&917.5,89.2&3.5$\pm$0.34&0.5$\pm$0.34&--&54$\pm$18&--&1400\\
         \hline
    \end{tabular}
        \tablenotetext{a}{The coordinates of the bottom of these jets as seen in IRIS 1400\,\AA.}
        \tablenotetext{b}{The errors are estimated from the instrumental resolution.}
    \tablenotetext{c}{The speed of the large jet is obtained from the S-T map of IRIS SJ 1400\,\AA, and that of a micro-jet is estimated by their lengths divided by the time it reaches the top.}
    \tablenotetext{d}{In what passbands the jet was clearly seen. EUV: the AIA EUV passbands; 1400: the IRIS 1400\,\AA\ passband.}
\end{table*}

\subsection{Properties of the large jet} \label{sec:floats}
\begin{figure*}
\includegraphics[trim=0cm 1cm 0cm 0cm, clip, width=\textwidth]{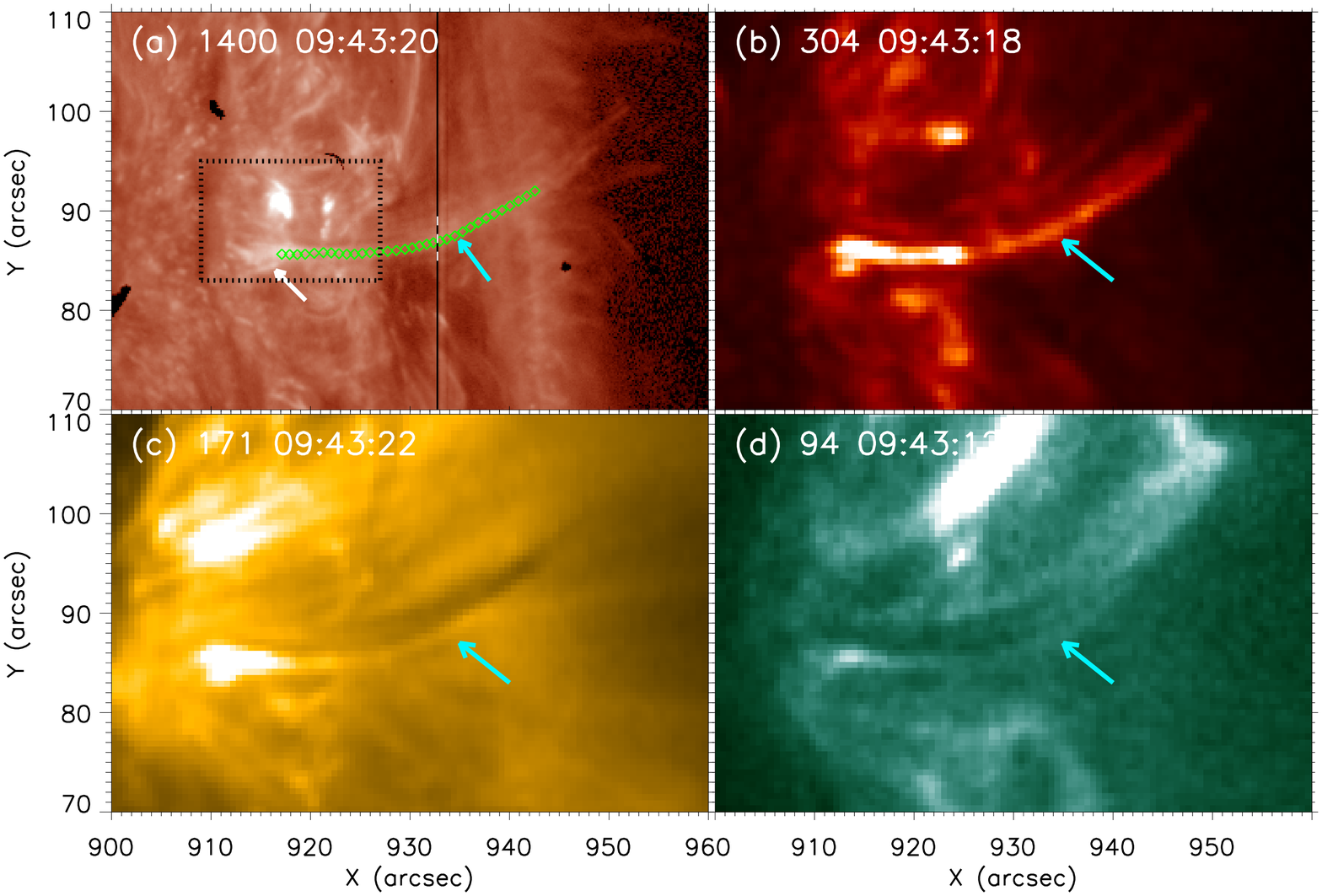}
\caption{The region of interest observed in the IRIS SJ 1400\,\AA\ (a) and AIA 304\,\AA\ (b), AIA 171\,\AA\ (c), AIA 94\,\AA\ (d) passbands.
The black dotted lines in panel (a) outline the field-of-view shown in Figure\,\ref{fig:jetfoot}.
The white arrow in panel (a) denotes the dome-like feature.
The green diamond symbols in panel (a) outline the trajectory of the EUV jet, which is also marked by the cyan arrows in the others.
The black vertical line in panel (a) is the location where the slit of the IRIS spectrograph targets on,
and the white dashed line indicates the spatial range where the slit images are shown in Figure\,\ref{fig:sp}\ and Figure\,\ref{fig:den}.
An associated animation is given online.
The animation includes evolution of this field-of-view seen in the four passbands from 09:16:39 UT to 10:28:55 UT.
\label{fig:overview}}
\end{figure*}

The large jet starts at 09:38 UT and lasts for about 10 minutes.
It has a curved trajectory, implying that it may flow along a large magnetic loop.
The jet can be seen in all the IRIS SJ and AIA EUV passbands (see the cyan arrows in Figure \ref{fig:overview}, where not all passbands are shown).
The brightenings at the base of the jet as viewed in these passbands peak at almost the same time and thus we can not distinguish which one has its first response.
In SJ 1400\,\AA\ passband, it consists of a compact bright thread wrapped by blurred surroundings (see the animation associated with Figure\,\ref{fig:overview}.)
We can see that it contains both bright and dark components in AIA 171\,\AA\ and 94\,\AA\ passbands (Figure\,\ref{fig:overview}(c) and Figure\,\ref{fig:overview}(d) for example).
The dark component of the jet shows up as bright structure in AIA 304\,\AA\ image while it is dark in the AIA channels with higher representative temperatures (see Figure\,\ref{fig:overview}).
While we also found the jet is associated with eruption of a twisted transition region loop (see section \ref{subsect_twflux}),
the dark component of the large jet is made of plasma which belonged to the transition region. 
Based on the response temperatures of the AIA passbands\,\citep{2010A&A...521A..21O}, 
this jet should have a multi-thermal nature including plasma with temperatures from transition region to corona.
Based on the SJ\,1400\,\AA\ images, its length is measured as 29\,\arcsec and the width of its compact bright component is about 1.5\,\arcsec.
The dark component best seen in the AIA 171\,\AA\ channel has a width of about 2.5\arcsec.


\begin{figure*}
\includegraphics[trim=0cm 5cm 0cm 2cm, clip,width=\textwidth]{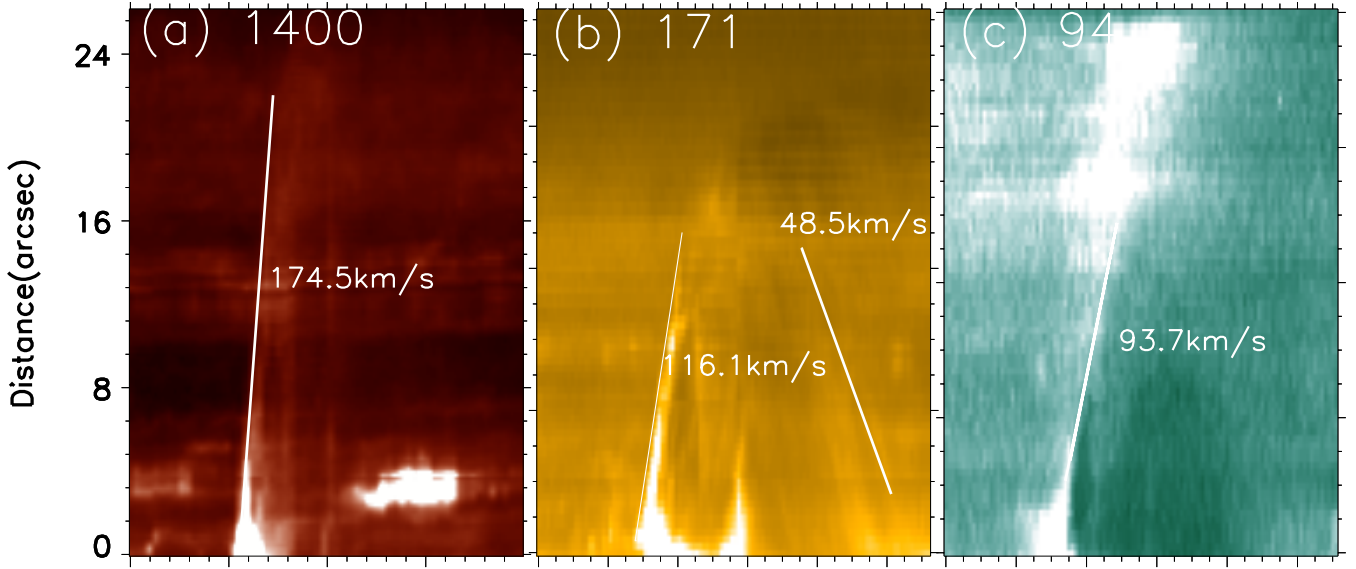}
\caption{Space-time plots along the trajectory of the EUV jet (see the green diamond symbols in Figure \ref{fig:overview}(a)) taken from IRIS SJ 1400\,\AA\ images (a), AIA 171\AA\ (b), and AIA 94\AA\ (c).
The apparent speeds derived from these plots are also marked.
\label{fig:stmap}}
\end{figure*}

\begin{figure*}
\includegraphics[width=\textwidth,trim=0cm 0cm 0cm 0cm]{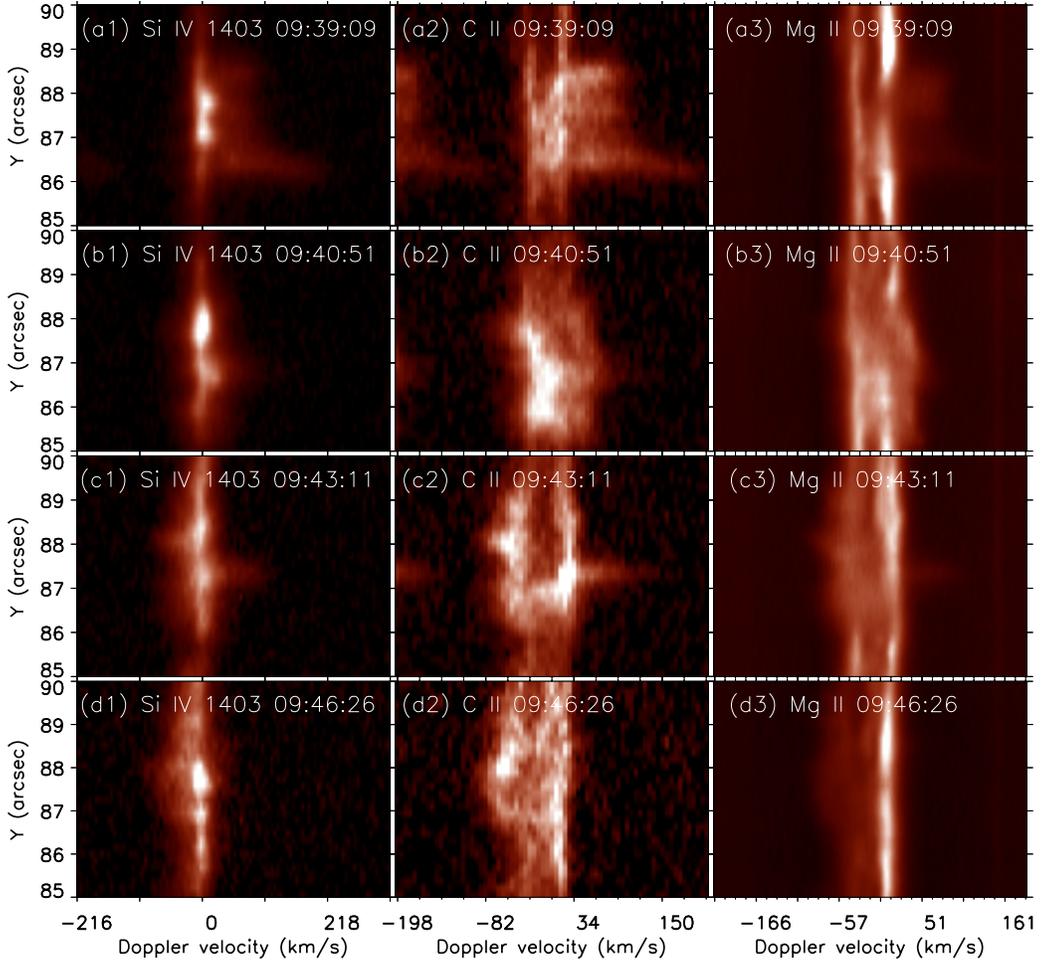}
\caption{Evolution of IRIS slit images of Si\,{\sc iv}\,1402.8\,\AA\ (1st column), C\,{\sc ii}\,1335.7\,\AA\ (2nd column) and Mg\,{\sc ii}\,2796.3\,\AA\ (3rd column) taken from the spatial range marked in Figure\,\ref{fig:overview}(a).
\label{fig:sp}}
\end{figure*}

\begin{figure*}
\includegraphics[trim=1cm 2.5cm 1cm 3cm,width=\textwidth]{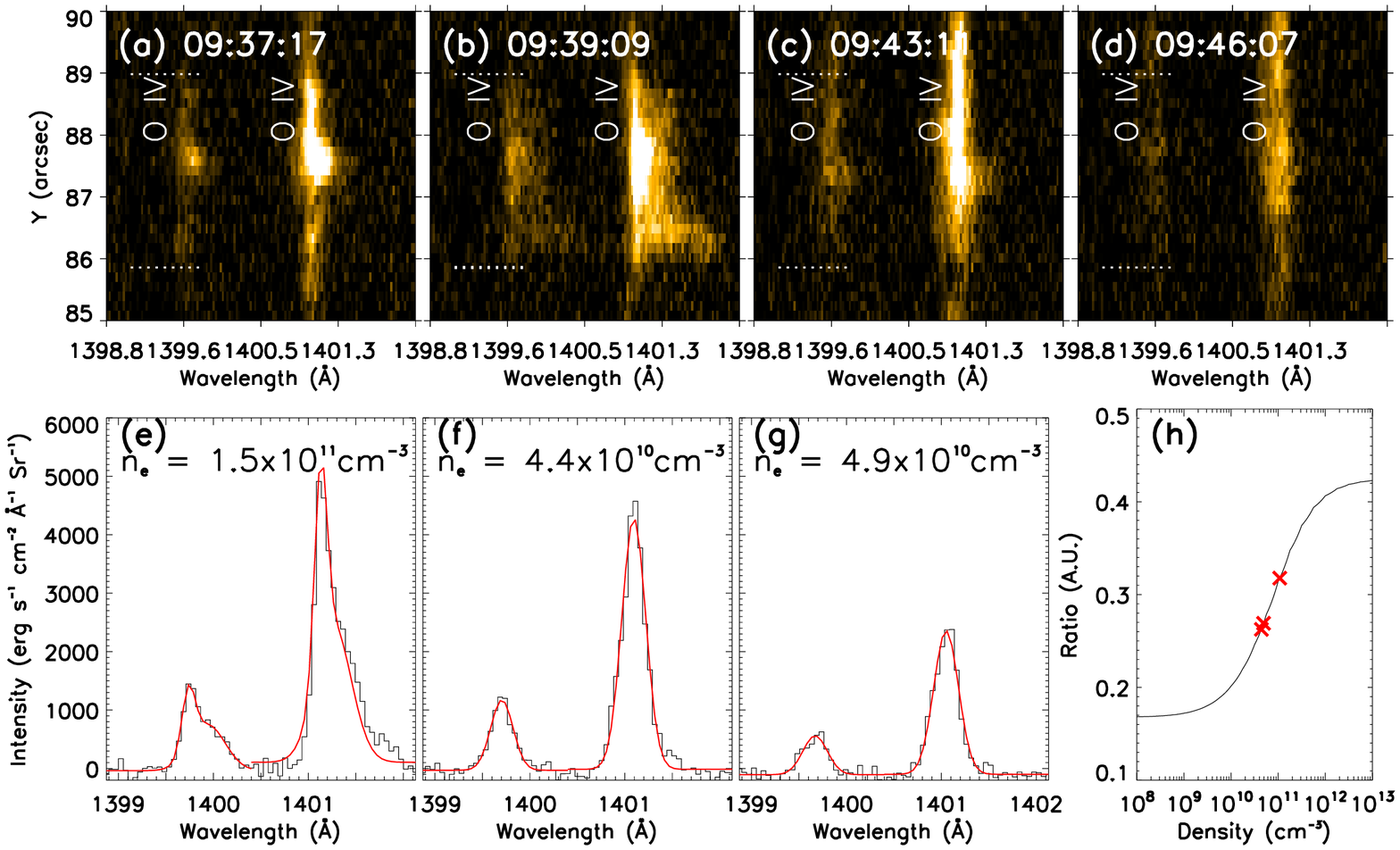}
\caption{Panels (a)-(d): IRIS slit images in the wavelengths around 1400\AA\ including the O\,{\sc iv} line pair of 1399.8\,\AA\ and 1401.2\,\AA\ at the location marked by white dash line in Figure \ref{fig:overview}(a) at four different times.
Panels (e)-(g): O\,{\sc iv} spectral profiles averaged cross the EUV jet (the region between the two horizontal dotted lines in panels (a)--(d)) 
at the times shown in panels (b),(c),(d),
in which the black lines are the observations and the red lines are the results from single or double Gaussian fits.
Panel (h): theoretical profile showing how the line ratio between the O IV 1399.8\AA\ and O IV 1401.2\AA\ varies with the electron densities given by CHIANTI,
and the line ratios from the cases in panel (e)-(g) are marked by red crosses and the derived electron densities are marked in panels (e)--(g).
\label{fig:den}}
\end{figure*}

\par
The large jet possesses both upward motion in early stage (as seen in all EUV passbands of AIA and IRIS SJIs) and downward motion in later stage (in all AIA EUV channels except 94\,\AA).
In Figure\,\ref{fig:stmap}, we show the space-time (S-T) maps obtained along the large jet as seen in IRIS SJ 1400\,\AA\ and AIA\,171\,\AA\ and 94\,\AA\ passbands.
Based on these S-T maps, we found that the upward motions have a velocity of $\sim$175 km/s in SJI 1400\,\AA, $\sim$116 km/s in AIA 171\,\AA\ and $\sim$94 km/s in AIA 94\AA, and the downward motions seen in AIA 171\,\AA\ have a velocity of $\sim$49 km/s (see Figure \ref{fig:stmap}).

\par
The spectrograph slit is perpendicular to the trajectory of the jet (see Figure\,\ref{fig:overview}(a)), which is about 15\arcsec\ away from the footpoint.
It allows us to study the evolution of the spectra of a cross-section of the jet.
In Figure\,\ref{fig:sp}, we show the evolution of the slit images of the spectral windows of Si\,{\sc iv}\,1403\,\AA, C\,{\sc ii}\,1336\,\AA\ and Mg\,{\sc ii}\,k\,2796\,\AA.
We can see that most spectra show clear enhanced wings.
The extensions of the enhanced wings of the spectra are different from one location to another,
demonstrating that the jet should consist of multiple fine threads.

\par
Focusing on the Si\,{\sc iv} spectra, one can see that non-Gaussian profiles are common.
While the jet is moving upward, we observe spectral profiles with enhanced wings at their red ends as far as at Doppler shifts of 150\,km/s (see Figure\,\ref{fig:sp} (a1)), again indicating the highly dynamic nature of the jet.
After that, the Doppler shifts of the enhanced red wings become smaller (see Figure\,\ref{fig:sp} (b1)).
At 09:43:11\,UT, about 5 minutes after the beginning of the jet, while the ejected materials start falling back, enhancements can be seen at both red and blue wings (Figure\,\ref{fig:sp} (c1)).
At 09:46:26\,UT, only blue-wing enhancements are seen (Figure\,\ref{fig:sp} (d1)) .
Such behaviours of the spectral profiles are consistent with the dynamics of the jet as seen by the IRIS and AIA imagers.
Such non-Gaussian profiles with strong emission in the line center and extensive enhanced wings are typical of the so-called ``explosive events''\,\citep[e.g.][]{1989SoPh..123...41D,1997Natur.386..811I,2014ApJ...797...88H,2017MNRAS.464.1753H}, which are thought to be signatures of magnetic reconnections via plasmoid instability\,\citep{2015ApJ...813...86I}.
We believe that these spectra in the present case are not due to reconnection because the location where the spectra are obtained is far away from the energetic sites. 
After carefully checking, we found that there exists a brightening in the background before arrival of the jet (Figure\,\ref{fig:den}(a)), and thus we believe that the strong emissions in the line cores are mostly from the background and the extended wings are due to the jet.

\par
With the IRIS spectral data of the O\,{\sc iv} line pair at 1399.8 and 1401.2\,\AA\ and the CHIANTI atomic database\,\citep{1997A&AS..125..149D,2019ApJS..241...22D}, we are also able to derive the electron density of the jet.
When the jet passes through the field-of-view of the slit (Figure\,\ref{fig:den}(b)--(d)), we produce the spectral profiles averaged over the cross-section of the jet and shown in Figure\,\ref{fig:den}(e)--(g).
The profile obtained at 09:39:09\,UT (at the stage of moving upward) is obviously non-Gaussian and can be fitted by a double Gaussian model (Figure\,\ref{fig:den}(e)).
Because of the dynamics of the jet, we believe the emissions at the wings are contributed by the jet.
With the line ratio using only the Gaussian components at the wings, the electron density is found to be $1.5\times10^{11}$\,cm$^{-3}$.
The mean profiles obtained at 09:43:11\,UT (about at the fall back) and 09:46:07\,UT (at the stage of falling back) are single-Gaussian as shown in Figure\,\ref{fig:den}(f) and Figure\,\ref{fig:den}(g).
The electron densities derived for those two line pairs are found to be $4.4\times10^{10}$\,cm$^{-3}$ and $4.9\times10^{10}$\,cm$^{-3}$, respectively.
The decrease of electron density is possible due to material falling back.
These values suggest that transition region materials have been ejected from the footpoint.

\begin{figure*}
\includegraphics[width=\textwidth,clip,trim=1cm 6.5cm 0cm 3.3cm]{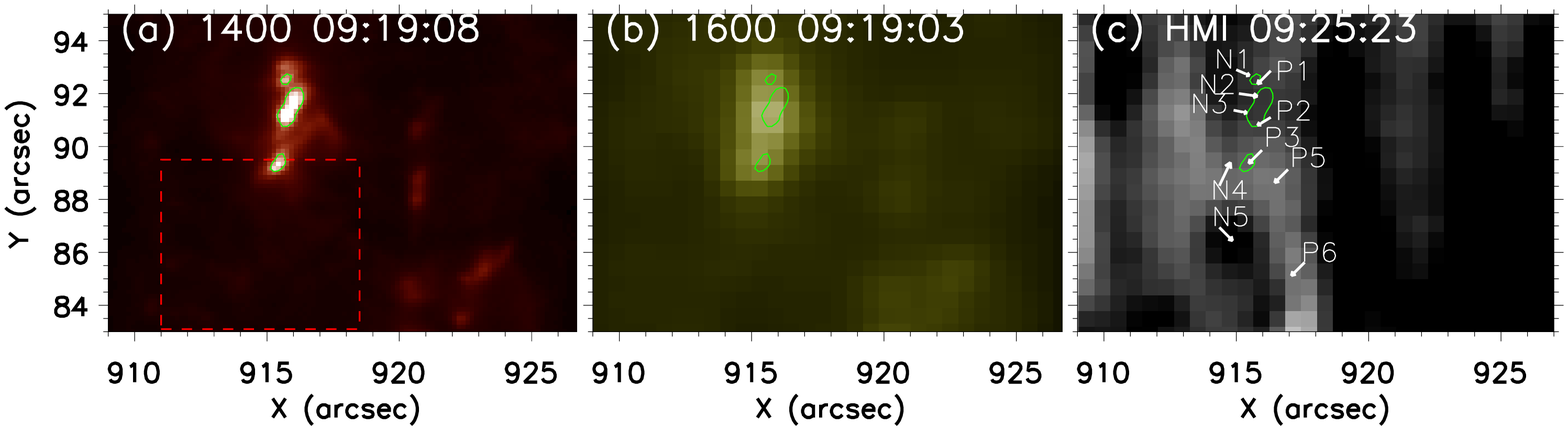}
\caption{A closer look at the region marked with black dotted square in Figure \ref{fig:overview}(a) taken with IRIS 1400\AA, AIA 1600\AA\ and the HMI line-of-sight magnetogram.
The green lines are contours at a level of 8000\,DN of the IRIS 1400\AA\ image.
The HMI magnetogram is scaled from --250\,G to 250\,G.
In panel (a), the dashed lines roughly enclose the region of the dome of the large jet.
In panel (c), the associated polarities of the brightening in IRIS 1400\AA\ image are denoted by the arrows with notes of  ``P1'',``P2'',``P3'' and ``N1'',``N2'',``N3'', which have magnetic strengths around the sensitive limit of the instrument and are mostly determined from the connectivities of the micro-loops as described in the main text. The associated polarities of the base of the EUV jet are marked as ``P5'',``P6'' and ``N5'', and ``N4'' marks the footpoint of an ``open'' field as determined from the first micro-jet (see the main text for detail).
\label{fig:jetfoot}}
\end{figure*}

\begin{figure*}
\includegraphics[width=\textwidth]{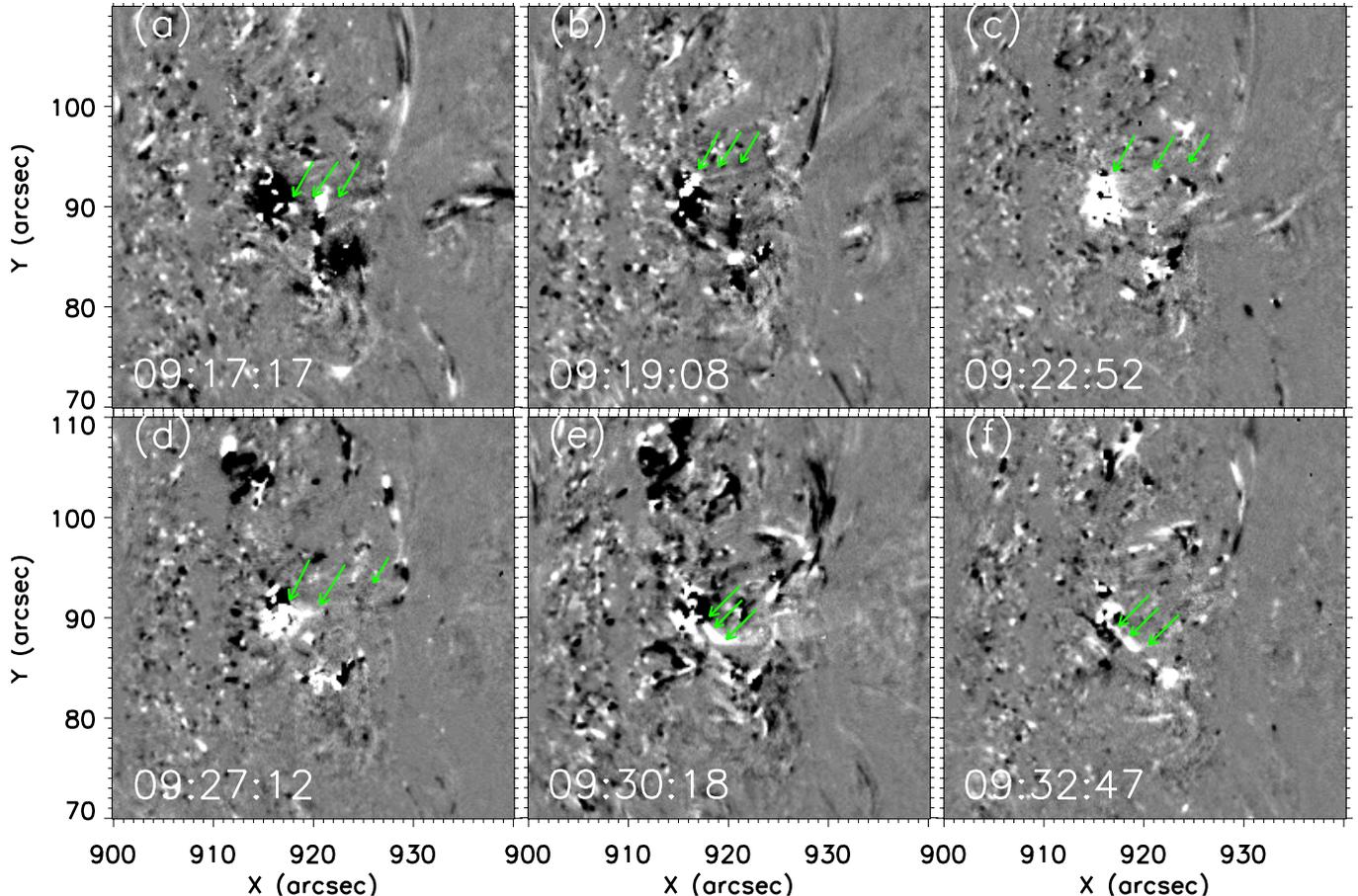}
\caption{Running difference images taken at the time when the micro-jets (denoted by green arrows that point to the bottom, middle and top parts of a jet) are clearly seen.
An associated animation is given online.
The animation includes evolution of this field-of-view from 09:17:17 UT to 09:48:55 UT.
\label{fig:mjs}}
\end{figure*}

\begin{figure*}
\includegraphics[width=0.9\textwidth,trim=1cm 3cm 1cm 2cm]{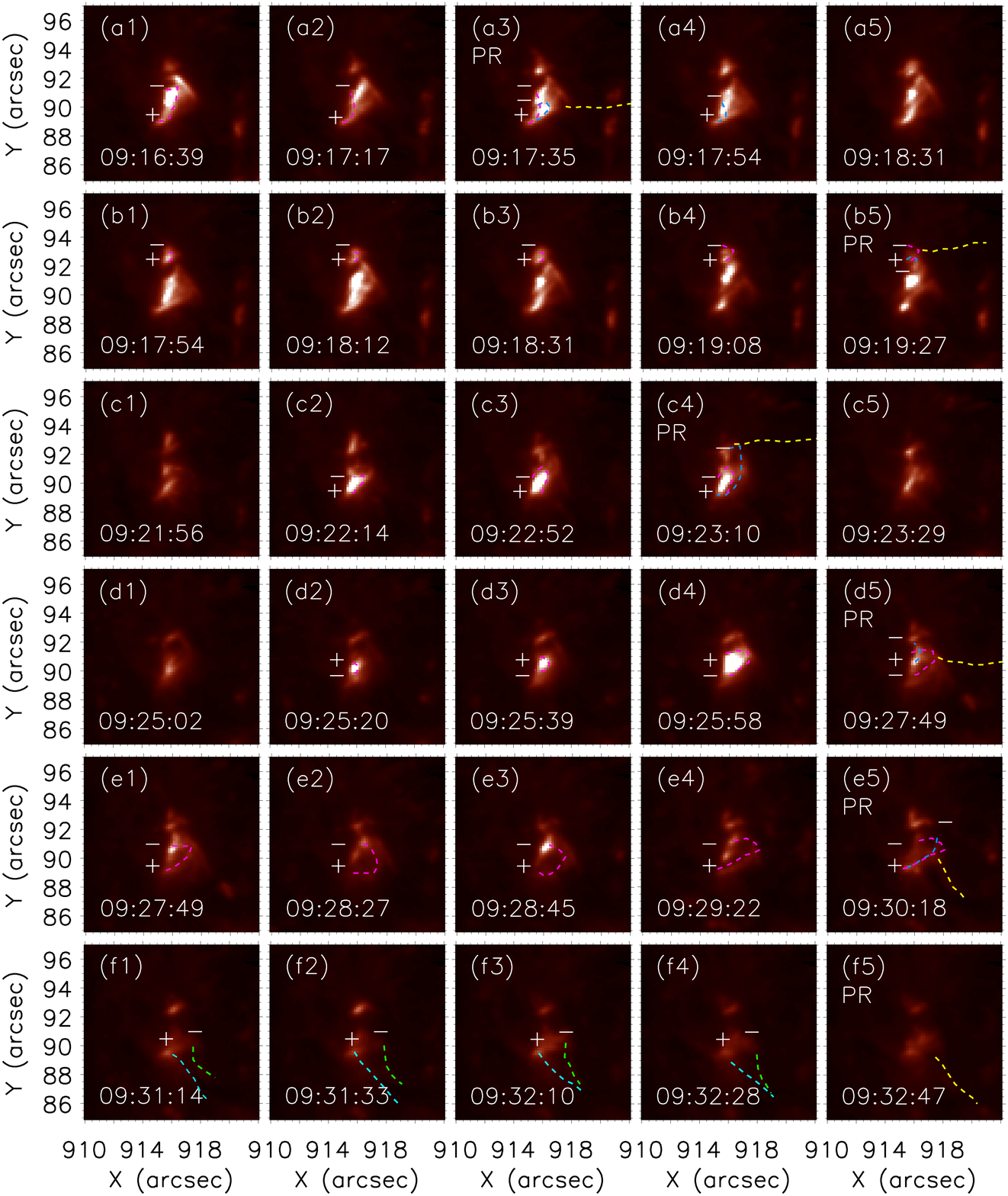}
\caption{Evolution of transition region micro-loops in the footpoint of the micro-jets as observed in the IRIS 1400\,\AA.
From top to bottom, each row corresponds to one micro-jet, in which (except the last row) the gold lines outline the micro-jet and the purple lines outline micro-loops seen before the micro-jet and the cyan lines outline that after.
A frame marked by ``PR'' is taken at the time just after occurrence of a micro-jet.
The cyan and green lines in the last row outline two elongated features that encounter each other and produce the sixth micro-jet.
The `+' and `--' symbols denote the positive and negative polarities, respectively.
An associated animation is given online.
The animation includes evolution of this field-of-view from 09:16:39 UT to 09:35:16 UT.
\label{fig:mlps}}
\end{figure*}

\subsection{Six transition region micro-jets} \label{subsec:microjets}
In Figure\,\ref{fig:jetfoot}, we show the SJ 1400\,\AA\ and AIA 1600\,\AA\ images and HMI magnetogram of the region around the footpoint of the large jet.
Six micro-jets are associated with a set of dynamic bright dots (BDs) found in the north of the dome-like structure (see Figure\,\ref{fig:jetfoot}(a)).
These BDs and micro-jets are clearly seen in IRIS SJ 1400\,\AA\ and AIA 1600\,\AA, but invisible in all AIA EUV channels (see Figures\,\ref{fig:overview}\ and \ref{fig:jetfoot}).
Therefore, we believe they are transition region structures.
These BDs are associated with micro loop-like features (hereafter, micro-loops) in the transition region, which will be described in details later.
Please note that these micro-loops mostly are very small and can be better followed in the associated animations.
The speed of a micro-jet is estimated as its length divided by the time it takes to reach its top.

\par
Although the region is close to the limb and the magnetogram is blurred, it can still give a rough picture of the magnetic environment of the events.
The HMI magnetogram (Figure\,\ref{fig:jetfoot}(c)) shows complex mixed polarities in the region.
Together with the locations of these BDs, their associated micro-loops and micro-jets and the dome-like structure, we identify a set of polarities that are crucial in the eruption of the sequence of the events (see Figure\,\ref{fig:jetfoot}(c)).
In these polarities, N1, N2, N3, P1, P2 and P3 are associated with the BDs and micro-loops, N4 links to ``open field'' that is determined from the locations of micro-jets, and N5 and the surrounding positive polarities (marked as P5 and P6) are associated with the dome-like structure.

\par
In Figure\,\ref{fig:mjs}, we show the IRIS SJ 1400\,\AA\ running difference images obtained at the time when the micro-jets are best seen.
Their dynamic evolution is easier to follow in the associated animation.
These micro-jets are very small in size, with less than 1\arcsec\ in width and a few arcseconds in length.
The footpoints of these micro-jets are changing back and forth in the south-north direction that will be described in details later.
To investigate the evolution of the footpoints of these micro-jets and the associated micro-loops, we show IRIS SJ 1400\,\AA\ images of the region in Figure\,\ref{fig:mlps} and the associated animation.
Please note that the micro-jets are hardly seen in these images because the scaling has been adjusted to show micro-loops in their footpoints.
The details of the evolution of these micro-jets are given below.

\par
The first micro-jet is first seen at 09:17:17 UT, and lasts for about 90\,s (denoted by the arrows in Figure \ref{fig:mjs}(a)).
Its width and length are about 0.7\arcsec\ and 4.6\arcsec, respectively.
Its speed is estimated to be about 47\,km/s.
In Figure\,\ref{fig:mlps}(a1)--(a5) and the associated animation, we show the evolution of the footpoint region of this micro-jet.
We observe micro transition region loop in the footpoint changes its connection simultaneously when the micro-jet is occurring.
Before occurrence of the micro-jet, a micro-loop connects the polarities of P3 and N2 (see the dashed lines in purple in Figure\,\ref{fig:mlps}(a1)--(a2)).
After the micro-jet, we see a new micro-loop connecting the polarities of P3 and N4 (see the dashed lines in cyan in Figure\,\ref{fig:mlps}(a3)--(a4)), which then disappears (see the associated animation).
The new micro-loop is smaller than the pre-existed one apparently.
Considering the location of the micro-jet (see the yellow line in Figure\,\ref{fig:mlps}(a3)), we believe that this micro-jet results from the reconnection between the micro-loop and an ambient open field.

\par
The second micro jet starts at about 09:19:08 UT (denoted by the arrows in Figure \ref{fig:mjs}(b)).
It can only be seen in three frames of the series of SJ 1400\,\AA\ images, and thus should have a lifetime less than 54\,s.
It has about 0.8\arcsec\ width and about 3.3\arcsec\ length.
Its speed is estimated to be 59\,km/s.
It has a typical inverted-Y shape that is better seen in the animation.
The connection of micro-loops in its footpoint changes from P1 and N1 before occurrence of the micro-jet to P1 and N2 after (see Figure\,\ref{fig:mlps}(b1)--(b5)).
 
\par
The third micro-jet starts at about 09:22:39 UT, and lasts for about 90\,s (denoted by the arrows in Figure \ref{fig:mjs}c).
It has about 0.7\arcsec\ width and 6.3\arcsec\ length.
Its speed is estimated to be about 50\,km/s.
Before occurrence of this micro-jet, we find that a micro-loop that connects the polarities of P2 and N2 moves toward north-west (see Figure\,\ref{fig:mlps}(c1)--(c3)).
While the micro-jet is clearly seen, a new micro-loop connecting the south footpoint of the previous micro-loop and that of the jet is formed (see Figure\,\ref{fig:mlps}(c4)--(c5)).
The newly formed micro-loop connects the polarities of P2 and N1, and it seems to be larger than the pre-existed one.

\par
The forth micro-jet starts at about 09:26:39 UT, and lasts for about 90\,s (marked by the arrow in Figure \ref{fig:mjs}(d)).
It has about 0.6\arcsec\ width and 8.0\arcsec\ length.
Its speed is estimated to be about 64\,km/s.
Before occurrence of this micro-jet, we observe that a brightening with a semi-circle shape appears in the footpoint, and we suggest it is a newly-emerged micro-loop.
While this brightening is growing, the micro-jet takes place at its north footpoint.
A micro-loop much smaller than the previous one is found in the footpoint of the micro-jet.

\par
The fifth micro-jet starts at about 09:30:15 UT, and lasts for about 144\,s (see the arrows in Figure \ref{fig:mjs}(e)).
It has 0.6\arcsec\ width and 3.0\arcsec\ length.
Its speed is estimated to be about 67\,km/s. 
Before occurrence of this micro-jet, we observe a rising micro-loop that connects the polarities of P3 and N3 (Figure\,\ref{fig:mlps}(e1)--(e4)).
The micro-jet apparently sets out from the top of the micro-loop during its expansion.
The size of micro-loops found after the micro-jet does not change much from that before.

\par
The sixth micro-jet starts at about 09:32:39 UT, and lasts for about 54\,s (see Figure \ref{fig:mjs}(f)).
It seems to be the consequence of the fifth micro-jet and its speed cannot be estimated.
We see the fifth micro-jet swings in the southeast direction.
It then attaches to an elongated feature that extends almost in parallel to the fifth micro-jet, and produces the sixth micro-jet (see Figure\,\ref{fig:mlps}(f1)--(f5)).
This micro-jet has about 0.5\arcsec\ width and 3.5\arcsec\ length.

\par
After eruptions of these micro-jets, at the north to the dome-like feature we observed twisted loops start untwisting and rising.
The rising of these twisted loops seem to trigger the large jet and lighten the dome-like structure.
Since these twisted loops can only be seen in the IRIS SJ 1400\,\AA\ passband, here we called them ``transition region flux ropes''.
Please note that we cannot confirm whether they are actually typical flux ropes because of the resolution of the instrument.
In the next section, we will describe in detail the evolution of the transition region flux ropes and the process that triggers the large jet from the dome-like structure.

\begin{figure*}
\includegraphics[width=\textwidth,trim=0.5cm 5.5cm 0cm 9.6cm,clip]{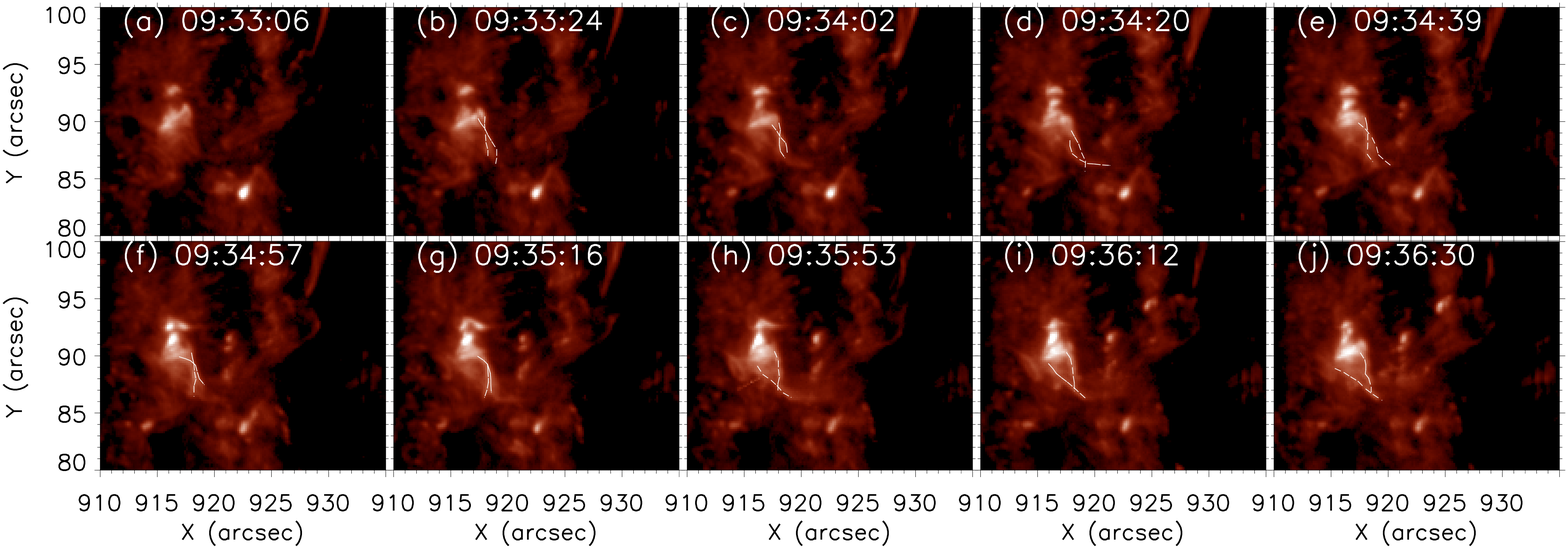}
\caption{Evolution of the transition region flux ropes that undergo rising and untwisting motions. The white dashed lines outline the identified threads of the flux ropes.
An associated animation is given online.
The animation includes evolution of this field-of-view from 09:31:33 UT to 09:48:17 UT.
\label{fig:twflux}}
\end{figure*}

\begin{figure*}
\includegraphics[width=\textwidth,trim=1.5cm 2.5cm 1cm 1cm,clip]{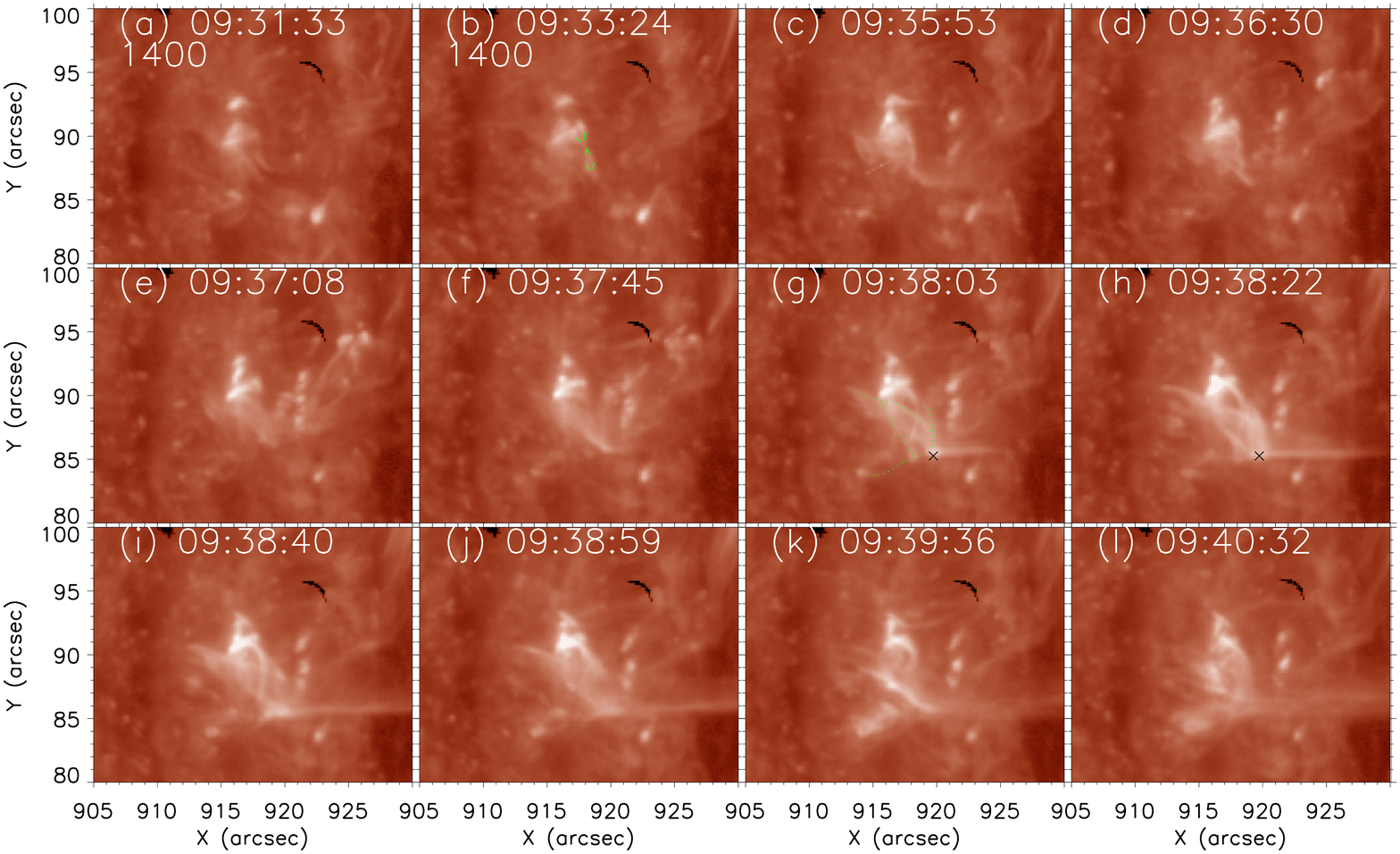}
\caption{Evolution of the footpoint region of the large jet.
The green dash line in panel (b) marks the rising twisted flux ropes as shown in Figure \ref{fig:twflux}.
The green dotted lines in panel (g) mark a few brightening threads in the base of the jet.
The black crosses in panel (g)-(h) mark the compact brightening that occurs at the start of the large jet.
\label{fig:jettrig}}
\end{figure*}

\subsection{Transition region flux ropes and triggers of the large jet}
\label{subsect_twflux}
The transition region flux ropes firstly appear in IRIS SJ 1400\,\AA\ images at about 09:32:10\,UT.
They extend toward southwest and connect a location near the footpoint of the fifth micro-jet and the south of the dome-like structure (see Figure\,\ref{fig:twflux} and the associated animation).
In the flux ropes, at least two bright threads can be identified in the SJ 1400\,\AA\ image (see the dashed lines marked in Figure\,\ref{fig:twflux}).
In Figure\,\ref{fig:twflux}(b)--(j), we outline the geometries of the two bright threads of the flux ropes when they are evolving.
By following their evolution , we can see untwisting and rising motions of these bright threads (better seen in the animation associated with Figure\,\ref{fig:twflux}).

\par
The triggering processes of the large jet can be seen in Figure\,\ref{fig:jettrig}.
At about 09:38:03\,UT, almost at the same time the large jet takes place, the dome-like structure is lightened up (see Figure\,\ref{fig:jettrig}(g)).
At the top of the dome, a compact brightening appears slightly before the brightening of the dome-like structure (see the cross symbols in Figure\,\ref{fig:jettrig}(g)--(h)), and that might be the magnetic reconnection point (or null point) of the large jet.
The brightenings in different fans of the dome-like structure appear successively.
The size of the base of the dome is about 8\arcsec\ in diameter (see Figure\,\ref{fig:jettrig}(g)--(l)).

\begin{figure*}
\includegraphics[width=\textwidth,clip,trim=0cm 0cm 0cm 0cm]{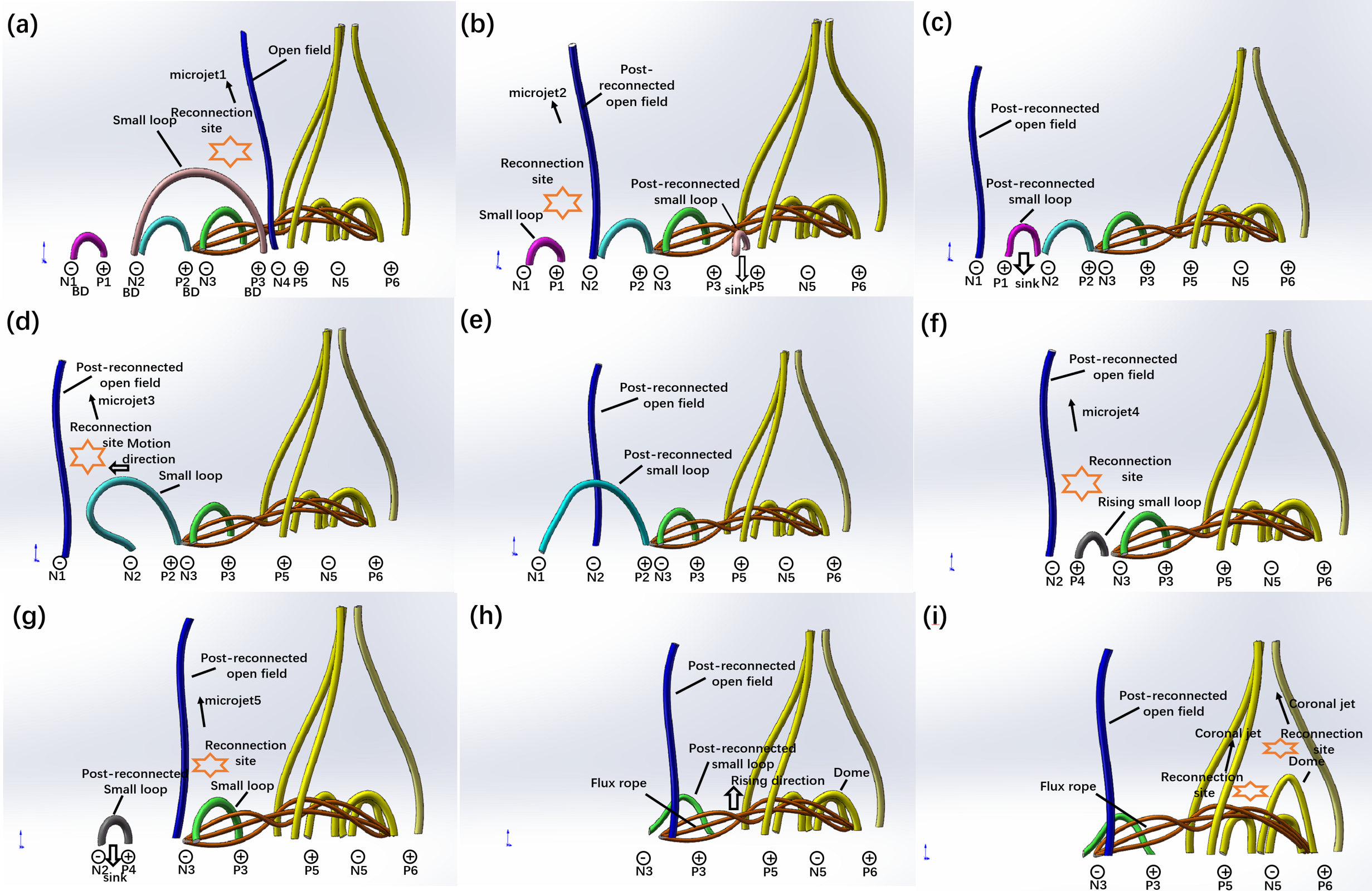}
\caption{A magnetic reconnection scenario interpreting the full set of the events. (See the main text for details.)
\label{fig:model}}
\end{figure*}

\section{Discussion}
\label{sect:dis}
Based on the observations, in Figure\,\ref{fig:model} we give a cartoon diagram to illustrate how the large jet may be triggered and what roles the series of TR micro-jets may play.
Before start of the sequence of events, the possible magnetic geometry is given in Figure\,\ref{fig:model}(a) where the polarities found in Figure\,\ref{fig:jetfoot} are also denoted.
Please notice that a polarity marked here is not infinitesimal, and field lines rooted in the same polarity as shown in the cartoon are not necessarily connecting the same point.
The bright dots (BDs) seen in the IRIS SJ 1400\,\AA\ are associated with the loops shown on the left; the dome-like structure is shown in gold on the right; and the transition region (TR) flux ropes are shown in brown and connect the system of BDs and the dome-like structure.
The TR flux ropes are suppressed by a loop system associated with the polarities of ``N2'', ``N3'', ``P2'' and ``P3'', and also possibly by the dome.
We assume the micro-jets flow along ``open'' field lines and thus their initial locations provide tracers to the locations of ``open'' field lines (blue lines in Figure\,\ref{fig:model}).

\par
The speeds of the micro-jets are relatively large in the transition region (compared to the sound speed and Alfv\'en speed in general).
We suggest these micro-jets are results of magnetic reconnection between the micro-loops and ``open'' field lines.
Taking into account the variation in the locations of these micro-jets and their concomitant dynamic micro-loops,
we further discuss the physical processes of these events in what follows.

\par
The sequence of the events starts from the first micro-jet, 
which results from magnetic reconnection between the micro loop which connects ``N2'' and ``P3'' (pink line in Figure\,\ref{fig:model}(a)) and nearby open field (blue line in Figure\,\ref{fig:model}(a) rooted in the polarity of ``N4'').
This step removes the first shell that constrains the TR flux ropes.
After this reconnection, the root of ``open'' field changes to ``N2'', and newly-formed micro-loop connecting ``N4'' and ``P3'' might submerge eventually.

\par
The second micro-jet takes place due to reconnection between the micro-loop connecting the polarities of ``N1'' and ``P1'' (purple line in Figure\,\ref{fig:model}(b)) and the ``open'' field rooted in ``N2''.
This step allows the ``open'' field to shift to ``N1'' and the newly-formed micro-loop to connect ``N2'' and ``P1''.
This newly-formed micro-loop sinks below the solar surface.
It allows the micro-loop connecting the polarities of ``N2'' and ``P2'' to rise (see the cyan line in Figure\,\ref{fig:model}(c)\ and Figure\,\ref{fig:model}(d).
This then triggers the third micro-jet and allows the ``open'' field shifts back to ``N2'' (see Figure\,\ref{fig:model}(e)).
The reconnection resulting in the third micro-jet actually generates a larger loop as demonstrated in the observations.
Please notice that Figure\,\ref{fig:model}(d) needs to be thought in three-dimension, and the reconnection is a typical component reconnection rather than an anti-parallel one.

\par
The forth micro jet appears to be result of reconnection between the ``open'' field and newly-emerging loops (see the grey line in Figure\,\ref{fig:model}(f)).
In this process, the root of the ``open'' field shifts to ``N3'', where close to the loop system that constrains the TR flux ropes (Figure\,\ref{fig:model}(g)).

\par
The fifth micro-jet could be result of another component reconnection between the ``open'' field and the micro-loop (see Figure\,\ref{fig:model}(g).
Although such reconnection might not change the general geometry of the magnetic field much,
the footpoint of the close loop switches to the location where the ``open'' field previously was, and that removes the last constraint on the TR flux ropes (see Figure\,\ref{fig:model}(h)).

\par
The TR flux ropes that have lost constraints might undergo untwisting and rising.
This can push the ``open'' field toward the dome and then cause occurrence of the sixth micro-jet.
The rising TR flux ropes can also push the fans in the dome upward.
The sixth micro-jet can also provide additional disturbance to the dome-like structure.
Under this circumstance, the large jet is triggered.
Both fans in the dome and the TR flux ropes can reconnect with the open fields of the dome (Figure\,\ref{fig:model}(i)). 
That generates the large jet, and it naturally contains both cool (in the TR flux ropes) and hot (heated in the reconnection) plasma.
At this stage, the eruption of the large jet follows the procedures as same as the mini-filament scenario\,\citep[e.g.][]{2015Natur.523..437S,2017Natur.544..452W}.

\par
Although the proposed scenario well explains the observations,
the current data do not include precise observations of magnetic field that is key to verifying any of these magnetic geometries.
This would require observations from a line-of-sight perpendicular to this region.
While these micro-jets might only be resolved from a side view, multiple satellites equipped with both magnetic and multi-wavelength imagers with high spatio-temporal resolution are essential to provide solid evidences for the scenario.
Such data might be achieved with multiple missions, such as a combination of Solar Orbiter\,\citep{2020A&A...642A...1M} and IRIS.
Future missions with multiple-perspective satellites, like the Solar Ring Mission\,\citep{2020ScChE..63.1699W}, if equipped with high-resolution EUV and magnetic imagers, will also help.
On the other hand, we cannot rule out the possibility that the TR flux ropes are not pre-existing but formed during occurrence of the series of TR micro-jets.

\section{Conclusions}
\label{sec:conc}
In the present paper, we report on IRIS and SDO observations of a sequences of jets in the solar atmosphere.
It consists of a large EUV jet and a series of transition region micro-jets prior to that.
We have conducted a detailed analysis of their physical properties and small-scale dynamics in their footpoints, including small-scale transition region (TR) loops and flux ropes and dome-like structure.
We found that the transition region micro-jets might play a crucial role in the eruption of the large jet.
Based on the observations, we propose a scenario to illustrate how the series of transition region micro-jets facilitate the large jet.

\par
The large jet originates from a dome-like structure and shows an inverted-Y geometry.
It includes both cool and hot components, with temperatures from that of the transition region to coronal.
It has about 29\arcsec\ in length, about 4\arcsec\ in width, and about 10 minutes of lifetime.
Its upward speed is higher than 170\,km/s as derived from the IRIS SJ 1400\,\AA\ space-time map, while its falling speed is much smaller ($\sim$50\,km/s).
Using the O\,{\sc iv} line pair of 1399.8\,\AA\ and 1401.2\,\AA, the electron density of the large jet is found to be about $1.5\times10^{11}$\,cm$^{-3}$.

\par
The series of micro-jets are visible only in the high resolution data from IRIS SJ 1400\,\AA\ passband, suggesting they are likely to be transition region phenomena.
They have sizes of $<$10\arcsec\ in length and $<$1\arcsec\ in width.
Their lifetimes are about 1--2 minutes and their speeds are in the range of 50--70\,km/s.
We find that these TR micro-jets are associated with dynamics of micro-loops (with length less than 10\arcsec) in the transition region.
Occurrence of a micro-jet is always accompanied by change of geometry of micro-loops in its footpoint,
and the location of the following micro-jet also changes accordingly.

\par
We find that a bundle of TR flux ropes connects the region of the micro-jets and the dome-like structure in the base of the large jet.
Following the micro-jets, the TR flux ropes undergo untwisting and rising motions and lead to eruption of the large jet and also lightening of the dome-like structure.
Our study is consistent with previous ones in that mini-filament eruption is the mechanism for a coronal jet,
and further suggests that such a ``mini-filament'' could be a transition region structure.

\par
Based on the observations, we propose a scenario involving a series of magnetic reconnection processes to explain the sequence of the events in the following steps:
(1) the TR flux ropes were constrained by micro-loops at the beginning;
(2) the constraints of the TR flux ropes are removed by magnetic reconnections that also generate the TR micro-jets;
(3) the TR flux ropes lose constraints and become unstable;
(4) the untwisting TR flux ropes interact with the dome, release their cool plasma into the spine of the dome and result in the large jet.

\par
Our observations demonstrate that the mass and energy couplings between the corona and lower atmosphere are complex.
To understand this complex process, small-scale dynamics in the lower solar atmosphere are crucial.
Although many of them might not carry directly mass and energy to the corona, they could be important for creating suitable magnetic and plasma environments that promote energetic events linking the solar corona and the lower atmosphere.

\begin{acknowledgments}
\textit{Acknowledgement}: We are grateful to the anonymous reviewer for his/her constructive and helpful comments and suggestions, and Prof. Bo Li for carefully reading the manuscript. 
This research is supported by National Key R\&D Program of China No. 2021YFA0718600 and National Natural Science Foundation of China (42174201, 41974201, 41874205, U1831112). 
IRIS is a NASA small explorer mission developed and operated by LMSAL with mission operations executed at NASA Ames Research center and major contributions to downlink communications funded by ESA and the Norwegian Space Centre.
The AIA and HMI data are used by courtesy of NASA/SDO, the AIA and HMI teams and JSOC.
We are grateful to Dr. Xiaoli Yan for helpful discussions and suggestions.
\end{acknowledgments}

\bibliography{jetbib}{}
\bibliographystyle{aasjournal}



\end{document}